\documentclass[preprint,showpacs,preprintnumbers,amsmath,amssymb]{revtex4} 
\usepackage{graphics}
\usepackage{dcolumn}
\usepackage{bm}
\usepackage{amsmath,amsthm,amscd,amssymb}
\usepackage{latexsym}
\usepackage{epsf}
\usepackage{epsfig}
\newcommand{\fr}{{^F\hspace{-.02in}R}}
\newcommand{\ed}{\operatorname{end}}
\newcommand{\de}{\operatorname{dec}}

\theoremstyle{plain}

\theoremstyle{definition}

\theoremstyle{remark}

\newtheorem{case[theorem]}{Case}

\begin{document}
\preprint{APS}
\title{A Gravitational Mechanism for the Acceleration of Ultrarelativistic Particles}
\author{C. Chicone}
\affiliation{Department of Mathematics\\University of
Missouri-Columbia\\Columbia, Missouri 65211, USA}
\author{B. Mashhoon}
\affiliation{Department of Physics and
Astronomy\\University of Missouri-Columbia\\Columbia, Missouri 65211, USA}
\date{\today }

\begin{abstract} 
Imagine a swarm of free particles near a point $P$ outside a 
gravitating mass $M$ and a free reference particle at $P$
 that is on a radial escape trajectory away from $M$. 
Relative to this reference particle and in a Fermi 
normal coordinate system constructed along its worldline, 
the particles in the swarm that move along the radial direction 
and are ultrarelativistic (that is, they have speeds above 
$c/\sqrt{2}$) 
decelerate toward this terminal speed. 
On the other hand, the swarm particles that are ultrarelativistic 
and move in directions normal to the radial (jet) directions 
accelerate to almost the speed of light by the gravitational tidal 
force of the mass $M$.
The implications of these effects as well as the influence of the 
higher-order terms on the tidal acceleration mechanism are investigated.
 The observational evidence in support of these 
general relativistic effects is briefly discussed.
\end{abstract}

\maketitle

\section{Introduction}
The geodesic deviation equation in general relativity is often compared with the Lorentz force law in electrodynamics.
Both connect the corresponding fields to the mechanics  of test particles and can thus be employed to provide
operational definitions of the gravitational and electromagnetic fields, respectively. In terms of the dynamics of {\em
ultrarelativistic} test particles, however, there is a fundamental difference between the two equations. To illustrate
this difference, let us consider the motion of a test particle of mass $m$ and charge $q$ in an electromagnetic field
$(\mathbf{E},\mathbf{B})$ in Minkowski spacetime with inertial coordinates 
$x^\mu =(t,\mathbf{x})$, 
\begin{equation}\label{eq1}
\frac{d}{dt}\left(\frac{\mathbf{v}}{\sqrt{1-v^2}}\right)=\frac{q}{m}(\mathbf{E}+\mathbf{v}\times \mathbf{B}),
\end{equation}
where $c=1$ throughout this paper. This equation can be written as
\begin{equation}\label{eq2}
\frac{d\mathbf{v}}{dt}=\frac{q}{m}\sqrt{1-v^2}\,[\mathbf{E}-(\mathbf{E}\cdot \mathbf{v})\mathbf{v}+\mathbf{v}\times \mathbf{B}].
\end{equation}
It is evident from this relation---as a result
of the appearance of $\sqrt{1-v^2}=1/\gamma$ as a factor on its
right-hand side---that as $v\to 1$, it becomes very difficult to 
change the velocity of the test particle; that is,
either the field should be maintained over a long integration time or 
the external electromagnetic field must become exceedingly 
strong. 
Otherwise, charged particles with $v^2$
 extremely close to unity simply behave as essentially 
free particles and travel along a straight line. 
This may be interpreted in terms of the energy of the 
ultrarelativistic particle, i.e. $d(m\gamma )/dt =q\mathbf{E}\cdot
\mathbf{v}$. For $v^2\to 1$, the particle has enormous energy and 
any significant change in its velocity produces a
significant change in the energy of the particle; therefore, unless an 
interaction with enormous energy is involved,
the particle can be considered to be essentially free. Moreover, in the limit of a massless particle such that $q/m $ 
remains finite as $m \to 0$, 
the particle simply follows a straight line at the speed of light, 
as expected. This situation should be contrasted with the well-known 
phenomenon of bending of a light ray in a gravitational field; 
therefore, the analogue of equation~\eqref{eq2} must be fundamentally 
different for the gravitational interaction. 
In fact, we will demonstrate that the analogue of the $1/\gamma$
 factor is absent in the gravitational case.
For one-dimensional motion along a straight line, \eqref{eq2} reduces to
\begin{equation}\label{eq3}
\frac{dv}{dt}=\frac{q}{m}(1-v^2)^{\frac{3}{2}}E_{\parallel},
\end{equation}
where $E_{\parallel}$ is the component of the electric field along the 
direction of motion of the particle.
 The relevant factor in this case is $1/\gamma^3$;
 again, we will show that such a factor is remarkably 
absent in the gravitational case.

Consider now the corresponding situation in a gravitational field. 
The analogue of equation \eqref{eq1} is the geodesic deviation equation. 
To express this equation in a form that can be compared with~\eqref{eq1}, 
we must establish a Fermi coordinate system about one of the 
geodesics~\cite{2}. In the curved spacetime of general
relativity, such a coordinate patch is the closest analogue of the global inertial system employed in equation
\eqref{eq1}. Let $X^\mu =(T,\mathbf{X})$ be the Fermi coordinate system that is valid in a cylindrical spacetime region
around the worldline of the reference geodesic observer $\mathcal{O}$. 
The motion of any other free test particle $\mathcal{A}$
in this neighborhood is given by the geodesic equation in Fermi coordinates
\begin{equation}\label{eq4}
\frac{d^2X^\mu}{ds^2}+\Gamma^\mu_{\alpha \beta}\frac{dX^\alpha}{ds}\frac{dX^\beta}{ds}=0,
\end{equation}
where $s$ is the proper time of $\mathcal{A}$ and $\Gamma^\mu_{\alpha \beta}$ 
is the spacetime connection in Fermi
coordinates. It is important to recognize that the
reference geodesic remains fixed throughout our analysis. This approach to
the deviation equation should be distinguished from the traditional one
based on  the Jacobi equation, which involves a certain linearization about
the reference geodesic. Rather, we simply solve the geodesic equation in a
Fermi normal coordinate system established along the reference geodesic.

Let us note that equations \eqref{eq1}--\eqref{eq3} are expressed with respect to the coordinate time $t$
of the global inertial system; therefore, it is necessary, for the purposes of comparison, to write equation \eqref{eq4}
in terms of the temporal Fermi coordinate $T$ as follows:
\begin{eqnarray}
\label{eq5} \frac{d^2T}{ds^2}&+&\Gamma^0_{\alpha \beta}\frac{dX^\alpha}{ds}\frac{dX^\beta}{ds}=0,\\
\label{eq6} \frac{d^2X^i}{ds^2}&+&\Gamma^i_{\alpha \beta}\frac{dX^\alpha}{ds}\frac{dX^\beta}{ds}=0.
\end{eqnarray}
Using the identity
\begin{equation}\label{eq7}
\frac{d^2X^i}{ds^2}=\frac{d^2T}{ds^2}\frac{dX^i}{dT}+\left( \frac{dT}{ds}\right)^2\frac{d^2X^i}{dT^2},
\end{equation}
equations~\eqref{eq5}--\eqref{eq7} imply that
\begin{equation}\label{eq8}
\frac{d^2X^i}{dT^2}+\left( \Gamma^i_{\alpha \beta}-\Gamma^0_{\alpha \beta} \frac{dX^i}{dT}\right) \frac{dX^\alpha }{dT} \frac{dX^\beta}{dT}=0.
\end{equation}
This is the analogue of equation~\eqref{eq1} in the theory of gravitation,
except that the geodesic worldline could in general be timelike,
null or spacelike. To ensure that we are dealing with  
a {\em timelike} worldline, 
 equation \eqref{eq4} must be supplemented with the requirement that 
\begin{equation}\label{eq9}
g_{\mu \nu} \frac{dX^\mu}{ds}\frac{dX^\nu}{ds}=-1,
\end{equation}
which is preserved by \eqref{eq4} throughout the motion. Writing the four-velocity of $\mathcal{A}$ in Fermi coordinates as
\begin{equation}\label{eq10}
U^\mu =\Gamma (1,\mathbf{V})
\end{equation}
with $\mathbf{V}=d\mathbf{X}/dT$, equation~\eqref{eq9} can be expressed as
\begin{equation}\label{eq11}
\Gamma^{-2}=-g_{00}-2g_{0i}V^i-g_{ij}V^iV^j.
\end{equation}
Here $\Gamma=dT/ds$ is the modified Lorentz factor of particle $\mathcal{A}$,
 so that $\Gamma \to \infty$ indicates that $ds\to 0$, 
i.e. the speed of particle $\mathcal{A}$ approaches the local speed of light 
and the worldline of $\mathcal{A}$ approaches a null geodesic.

The Christoffel symbols in equation \eqref{eq8} are obtained from the metric tensor in Fermi coordinates
\begin{eqnarray}
\label{eq12} g_{00}&=&-1-\fr_{0i0j} X^iX^j+O(|\mathbf{X}|^3),\\
\label{eq13} g_{0i}&=&-\frac{2}{3}\, {\fr_{0jik}}X^jX^k+O(|\mathbf{X}|^3),\\
\label{eq14} g_{ij}&=&\delta_{ij}-\frac{1}{3}\, {\fr_{ikjl}}X^kX^l
                       +O(|\mathbf{X}|^3),
\end{eqnarray}
where
\begin{equation}\label{eq15}
\fr_{\alpha \beta \gamma \delta}(T)=R_{\mu \nu \rho \sigma}\lambda^\mu_{\;\;(\alpha )}\lambda^\nu_{\;\;(\beta)} \lambda^\rho_{\;\;(\gamma)} \lambda^\sigma_{\;\; (\delta)}
\end{equation}
is the projection of the Riemann tensor on the orthonormal parallel-propagated 
tetrad frame $\lambda^\mu_{\;\;(\alpha)}$ along the worldline of $\mathcal{O}$. Here $\lambda^\mu_{\;\;(0)}=dx^\mu
/d\tau$ is the four-velocity vector of $\mathcal{O}$ and constitutes the temporal axis of its local frame while
$\lambda^\mu_{\;\;(i)}$, $i=1,2,3$, 
are the unit spatial axes that constitute the orthonormal spatial triad of the 
local frame such that
\begin{equation}\label{eq16}
g_{\mu \nu}\lambda^\mu_{\;\;(\alpha )}\lambda^\nu _{\;\;(\beta )}=\eta_{\alpha \beta}.
\end{equation}
Here $\eta_{\alpha \beta}$ is the Minkowski metric tensor with 
signature $+2$ and $\tau$ is the proper time of $\mathcal{O}$; 
moreover, the Fermi system is so constructed that for this reference observer 
$(T,\mathbf{X})=(\tau ,\mathbf{0})$.

The infinite series in \eqref{eq12}--\eqref{eq14} consist of
increasing powers of the relative distance with time-dependent 
coefficients that involve derivatives of the Riemann tensor along the worldline of $\mathcal{O}$. 
The general behavior of the higher-order terms has been discussed
in~\cite{5,n6}; in fact, though the higher-order terms in~\eqref{eq12}--\eqref{eq14}
can be computed in principle, this has actually been done only up to the terms of 
order $|\mathbf{X}|^4$. 
The Fermi coordinates are admissible for $|\mathbf{X}|<\mathcal{R}$, 
where $\mathcal{R}(T)$ is the radius of the cylindrical region along 
the worldline of $\mathcal{O}$.  To determine $\mathcal{R}$, 
consider the nonzero components of the
Riemann tensor and its covariant derivatives along the 
reference geodesic $\mathcal{O}$; then, $\mathcal{R}$ is defined to be 
\begin{equation}\label{radfc}
\inf\big\{\frac{1}{|\fr_{\alpha\beta\gamma\delta}|^{1/2}},\, 
\Big|\frac{\fr_{\alpha\beta\gamma\delta}}{\fr_{\alpha\beta\gamma\delta,\rho}}\Big|,\,
\Big|\frac{\fr_{\alpha\beta\gamma\delta,\rho}}{\fr_{\alpha\beta\gamma\delta,\rho\sigma}}\Big|,\ldots
\big\}.
\end{equation}
The relationship between $\mathcal{R}$ and the radii of
convergence of the series in~\eqref{eq12}--\eqref{eq14} is not known.

Using the explicit form of the metric tensor in \eqref{eq12}--\eqref{eq14}, equations \eqref{eq8} and \eqref{eq11} can be expressed as
\begin{widetext}\begin{eqnarray}\label{eq18}
&& \frac{d^2X^i}{dT^2} + \fr_{0i0j} X^j+2\,\fr_{ikj0}V^kX^j\notag\\
&&+\left( 2\,\fr_{0kj0} V^iV^k+\frac{2}{3}\, {\fr_{ikjl}}V^kV^l
  +\frac{2}{3}\,{\fr_{0kjl}}V^iV^kV^l\right) X^j+O(|\mathbf{X}|^2)=0,
\end{eqnarray}\end{widetext}
and
\begin{eqnarray}\label{eq19}
\frac{1}{\Gamma^2} &= & 1-V^2+\fr_{0i0j}X^iX^j+\frac{4}{3}\, 
\fr_{0jik} X^jV^iX^k\notag\\
& & \quad +\frac{1}{3}{\fr_{ikjl}}V^iX^kV^jX^l+O(|\mathbf{X}|^3)>0.
\end{eqnarray}
Equation~\eqref{eq18} should be compared and contrasted with 
equation~\eqref{eq2}: 
There is no impediment in the gravitational case for changing the 
velocity of an ultrarelativistic particle with $|\mathbf{V}|$
initially very 
near unity in contrast with electrodynamics. Moreover,
 it follows from equation \eqref{eq5} that in the gravitational case, 
the analogue of the electromagnetic particle energy equation takes the form
\begin{equation}\label{eq19a} 
\frac{1}{\Gamma}\frac{d\Gamma}{dT}
=-\Gamma^0_{\alpha \beta} \frac{dX^\alpha}{dT}\frac{dX^\beta}{dT},
\end{equation}
so that unlike the electrodynamic case, 
the rate of variation of particle energy is in fact proportional 
to the energy of the particle.
For one-dimensional relative motion in the $X$ direction with $V=dX/dT$, 
equation~\eqref{eq18} reduces to
\begin{equation}\label{eq20}
\frac{dV}{dT}+\kappa (1-2V^2)X+O(X^2)=0,
\end{equation}
where $\kappa=\fr_{TXTX} (T)$. Equation~\eqref{eq20} should be 
compared and contrasted with equation~\eqref{eq3}.
Neglecting higher-order terms in~\eqref{eq20}, we note the existence of a critical speed $V_c=1/\sqrt{2}\approx 0.7$ in the
gravitational case. The physics of this critical speed has been discussed in 
our previous 
work \cite{5,6, n4,n5,n6}; indeed, the existence of the critical speed 
is basically due to the fact
that the motion of $\mathcal{A}$ is expressed with respect to the Fermi
coordinate time $T$, which reduces to the proper time of the reference
observer $\mathcal{O}$ along its worldline.

Consider the solution of \eqref{eq20} with initial conditions that at 
$T=0$, $X=0$ and $V=\sqrt{\Gamma^2_0-1}/\Gamma_0$ such that $\Gamma_0>>1$.
 Ignoring higher-order tidal terms, a simple integration of \eqref{eq20} demonstrates that depending on the 
sign and magnitude of $\kappa$, one can raise or lower the velocity
$V$ considerably. In fact, $V$ can be raised to a value such that the quantity
$\Gamma^{-2}$ corresponding to equation~\eqref{eq19}, i.e.
\begin{equation}\label{eq21}
\Gamma^{-2}=1-V^2+\kappa X^2+O(X^3),
\end{equation}
vanishes provided that higher-order terms are neglected. 
This singular behavior signals the breakdown of the test 
particle approach adopted in our treatment, where higher-order tidal terms
are perforce ignored. 
These issues will be illustrated in the next section in the context of the 
simplest gravitational field of astrophysical interest, namely, 
the exterior Schwarzschild spacetime.
It turns out that the results are observationally significant only
near gravitationally collapsed configurations.

\section{Tidal Effects in Schwarzschild Spacetime}
Imagine a swarm of relativistic particles around
a point $P$ in the exterior Schwarzschild spacetime that
corresponds to the gravitational field of a mass $M$ embedded
in an otherwise flat spacetime with asymptotically inertial 
coordinates $ (t, x,y, z)$. Let the radial line joining the center of
the spherical symmetry to $P$ be the $z$ axis. Then in terms of the 
corresponding spherical polar coordinates $(r,\theta,\phi)$, the
Schwarzschild metric is 
\begin{equation}\label{sm}
-ds^2=-(1-\frac{2 G M}{r})\,dt^2+(1-\frac{2 G M}{r})^{-1}\,dr^2
+r^2(d\theta^2+\sin^2\theta\,d\phi^2).
\end{equation}
\begin{figure}
\centerline{\psfig{file=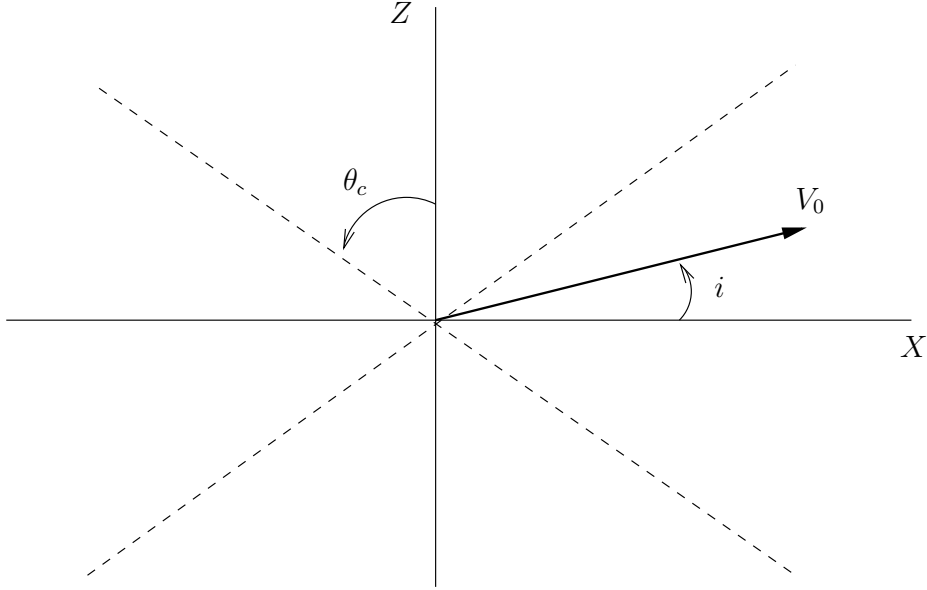, width=30pc}}
\caption{ Plot of the initial position and velocity of a particle
in the Fermi coordinate system. The dashed lines represent the critical
velocity cone demarcating the acceleration domains from the deceleration
domains. The critical angle $\theta_c$ is given by
$\tan\theta_c=1/V_c=\sqrt{2}\,$, so that $\theta_c$ is approximately
$54.7^{\,\circ}$. \label{fig:0}}
\end{figure}

It is convenient to refer the motion of the particles
in the swarm to an observer $\mathcal{O}$ that starts from $P$ at 
$r=r_0>2GM$ and moves relatively slowly along the radial $z$ direction reaching
infinity with a speed that is very small compared to unity. 
Our results turn out to
be essentially independent of such a speed; therefore, we set it equal to
zero for the sake of simplicity.
The geodesic path of $\mathcal{O}$ is thus given by
\begin{equation}\label{sdes}
\frac{dt}{d\tau}=(1-\frac{2 G M}{r})^{-1},\qquad \frac{dr}{d\tau}
=\sqrt{\frac{2GM}{r}}.
\end{equation}
We then set up a Fermi coordinate system $(T, X,Y,Z)$ about the 
worldline of $\mathcal{O}$ such that $\lambda^\mu_{\;\; (3)}$ is
along the radial direction.
In fact, in the $(t, r,\theta,\phi)$ coordinates we choose
\begin{equation}\label{neweq24}
\lambda^\mu_{\;\; (3)}= ( \sqrt{\frac{2 GM}{r}}\, 
\Big(1-\frac{2 GM}{r}\Big )^{-1},1,0,0).
\end{equation}
There is then a rotational degeneracy in the choice of  $\lambda^\mu_{\;\; (1)}$
and $\lambda^\mu_{\;\; (2)}$; once these vectors are chosen at a given 
instant of time, they are then parallel transported along the worldline.
The nonzero components of the Riemann tensor along the path of $\mathcal{O}$
are given by
\begin{eqnarray}\label{rtfl}
\fr_{0101}=\fr_{0202}=-\frac{1}{2} k,\qquad \fr_{0303}=k,\\
\label{rtf2} \fr_{2323}=\fr_{3131}=\frac{1}{2} k,\qquad \fr_{1212}=-k,
\end{eqnarray}
except for the symmetries of the Riemann tensor. Here $k$ is given by
 \begin{equation}\label{kdef}
k=-\frac{2GM}{r^3},
\end{equation}
which can be expressed via the integration of~\eqref{sdes} and $\tau\mapsto T$
as 
\begin{equation}\label{kdefitsT}
k(T)=-2GM(r_0^{3/2}+\frac{3}{2} \sqrt{2GM}\,T)^{-2}.
\end{equation}
More generally, the evaluation of equation \eqref{eq15} for an arbitrary timelike geodesic in Schwarzschild spacetime
has been considered in \cite{7}. 

The equations of motion of the particles in the swarm relative to $\mathcal{O}$
are thus
\begin{eqnarray}\label{swarmeqmx}
 \ddot X-\frac{1}{2}kX[1-2 \dot X^2+\frac{2}{3}(2\dot Y^2-\dot Z^2)]
+\frac{1}{3} k\dot X(5 Y \dot Y -7Z\dot Z)&=&0,\\
\label{swarmeqmy} \ddot Y-\frac{1}{2}kY[1-2 \dot Y^2+\frac{2}{3}(2\dot X^2-\dot Z^2)]
+\frac{1}{3} k\dot Y(5 X \dot X -7 Z\dot Z)&=&0,\\
\label{swarmeqmz}\ddot Z+kZ[1-2 \dot Z^2+\frac{1}{3}(\dot X^2+\dot Y^2)]
+\frac{2}{3} k\dot Z( X \dot X + Y\dot Y)&=&0,
\end{eqnarray}
together with the timelike condition, namely, that 
\begin{equation}\label{timecond}
\Gamma^{-2}=1-V^2-\frac{1}{2}k (X^2+Y^2-2 Z^2)
+\frac{1}{6}k[(\dot X Z-X\dot Z)^2+ (\dot Y Z-Y\dot Z)^2-2(\dot X Y-X \dot Y)^2]
\end{equation}
be positive. 
Here the overdot denotes differentiation with respect to the Fermi time
$T$; moreover, we have limited our attention to 
terms given explicitly in equations~\eqref{eq12}--\eqref{eq14}.
Equations~\eqref{swarmeqmx}--\eqref{swarmeqmz} and~\eqref{timecond}
can  be put in dimensionless form if all spatial
and temporal durations are expressed in units of $GM$.
There is therefore a simple scaling law at work here for
different mass scales (for instance, microquasars $\to$ quasars).
Our results are most important very near the source, since the tidal
forces decrease as $r^{-3}$ away from the source.

To simplify matters, we take advantage of the axial symmetry of this system
about the $Z$ direction and set $Y(T)=0$ for all $T$. 
The resulting system can be integrated with the initial conditions that 
at $T=0$,
$X=Z=0$ and 
\begin{equation}\label{xzcond}
\dot X=V_0\cos i,\qquad \dot Z=V_0\sin i,
\end{equation}
where $i$ is the inclination angle and $V_0$ is the initial speed
$V_0=\sqrt{\Gamma_0^2-1}\,/\Gamma_0$ such that $\Gamma_0\gg 1$, 
see Fig.~\ref{fig:0}. 
\begin{figure}
\centerline{\psfig{file=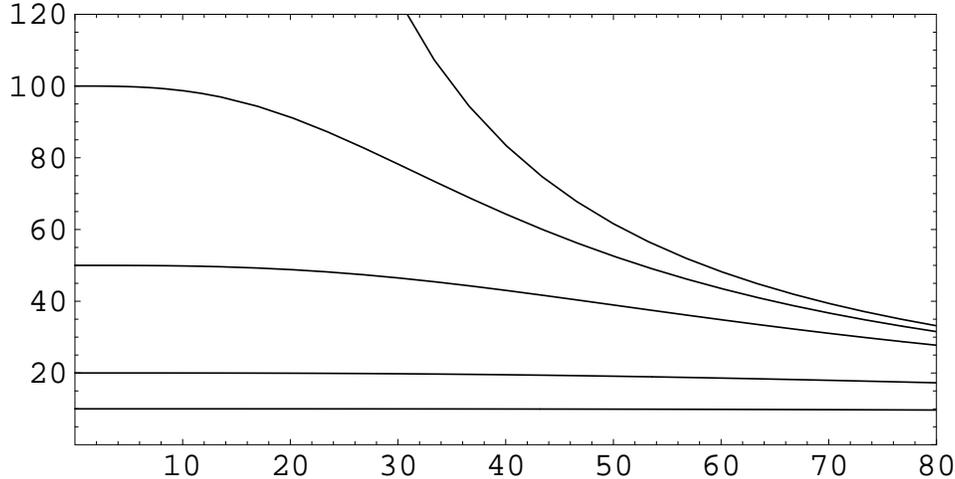, width=30pc}}
\caption{Plot of the Lorentz factor 
$\Gamma=1/\sqrt{1-\dot Z^2-2 GM Z^2/r^3}$ versus
$T/(GM)$
based on the integration of equation~\eqref{eq32}
for $r_0=100\,GM$ and $\Gamma_0=10, 20, 50, 100$  and 1000. 
\label{fig:1}}
\end{figure}

For the motion of a free test particle $\mathcal{A}$
 along the (radial) $Z$ direction, $i=\pi/2$, the equations of 
motion~\eqref{swarmeqmx}--\eqref{swarmeqmz}
reduce to the
nonlinear equation
\begin{equation}\label{eq32}
\ddot{Z}+kZ(1-2\dot{Z}^2)=0.
\end{equation}
The main aspects of
nonlinearity here
that are
significant for
our discussion are
the existence of
critical solutions
involving uniform
motion at $V_c=\pm
1/\sqrt{2}$ and the effective change
in the sign of curvature $k$ for $\dot Z^2>1/2$. For
initially ultrarelativistic motion (i.e. $V_0>1/\sqrt{2}$), 
the particle decelerates toward the terminal speed
$1/\sqrt{2}$ as demonstrated in Fig.~\ref{fig:1}. The details
of the deceleration process depend upon $r_0> 2\,GM$. However,
it is important to point out a \emph{general feature} of the tidal 
deceleration mechanism in the particular case under consideration
in Fig.~\ref{fig:1}: the particle starts at $T=0$ with
$r_0=100\, GM$, $Z=0$ and $\dot Z>1/\sqrt{2}$  
and decelerates to $\Gamma<40$ during a time $T\approx 80\, GM$ {\em regardless} of the initial 
$\Gamma_0\ge 50$. Thus, for $\Gamma_0\to\infty$, ``infinite'' deceleration
can occur in a relatively short period of time; however, in this 
limiting case the test particle approximation breaks down. 

On the other hand, for motion along the $X$ direction, $i=0$, 
the equations of motion reduce to an equation similar to
\eqref{eq32},
\begin{equation}\label{eq33}
\ddot{X}-\frac{1}{2}kX(1-2\dot{X}^2)=0,
\end{equation}
except that an initially ultrarelativistic particle accelerates 
along directions normal to the jet direction as
illustrated in Fig.~\ref{fig:2}. As before, 
the particle starts at $ T = 0$  with $r_0 = 100\, GM$, $X = 0$ and 
$\dot X > 1/\sqrt{2}$; 
the integration ends at $ T_{\ed} $ when $\Gamma \to\infty$.
We note that $T_{\ed}$ may occur outside the domain of validity
of Fermi coordinates; however, as $\Gamma_0 \to\infty$, 
$X_{\ed} = X (T_{\ed})\approx T_{\ed}$ tends to zero and one might expect that 
in this case the influence
of the higher-order terms may be small. It is in fact possible to show that
in this limit $T_{\ed}$ approaches zero as $\Gamma_0^{-2/3}$. This is demonstrated in
appendix A. It is necessary to remark here that for sufficiently
large $\Gamma_0$, the influence of the particle on the background geometry can
no longer be neglected and hence our treatment becomes invalid.

    To discuss the physics of the tidal acceleration/deceleration phenomena,
it is useful to have an invariant measure of the energy of the test particle
$\mathcal{A}$. Let us therefore introduce a class of static observers with
four-velocity $U_R^{\;\;\mu}$ in the Fermi coordinate system such that in the 
$(T, X, Y, Z)$ coordinates
\begin{equation}\label{neweq34}
U_R^{\;\;\mu}(T,\mathbf{X})=\Big(\frac{1}{\sqrt{-g_{00}}}, 0,0,0\Big).
\end{equation}
These fundamental observers associated with Fermi coordinates are in general
accelerated; of course, one exception is the reference geodesic $\mathcal{O}$
that is at rest at the spatial origin of the Fermi system. At any given
event along the path of $\mathcal{A}$, the corresponding 
fundamental observer at
that event measures the energy per unit mass of $\mathcal{A}$ to be 
$\hat\Gamma = - g_{\mu \nu} U_R^{\;\;\mu} U^\nu$, or
\begin{equation}\label{neweq35}
\hat\Gamma=\Big(\sqrt{-g_{00}}-\frac{g_{0i}V^i}{\sqrt{-g_{00}}}\Big) \Gamma.
\end{equation}
In the approximation scheme of this section, we have 
$\hat\Gamma= \sqrt{-g_{00}}\,\Gamma$, so that with 
\begin{equation}\label{g00}
-g_{00} = 1 + \frac{GM}{ r^3} ( X^2 + Y^2 - 2 Z^2 ),
\end{equation}
$\hat\Gamma$ decreases in the deceleration case ($X = Y = 0$) and
increases in the acceleration case ($ Y = Z = 0 $). Therefore, employing the
invariant $\hat\Gamma$ instead of $\Gamma$ would simply 
enhance these main results
of the present section.

\begin{figure}
\centerline{\psfig{file=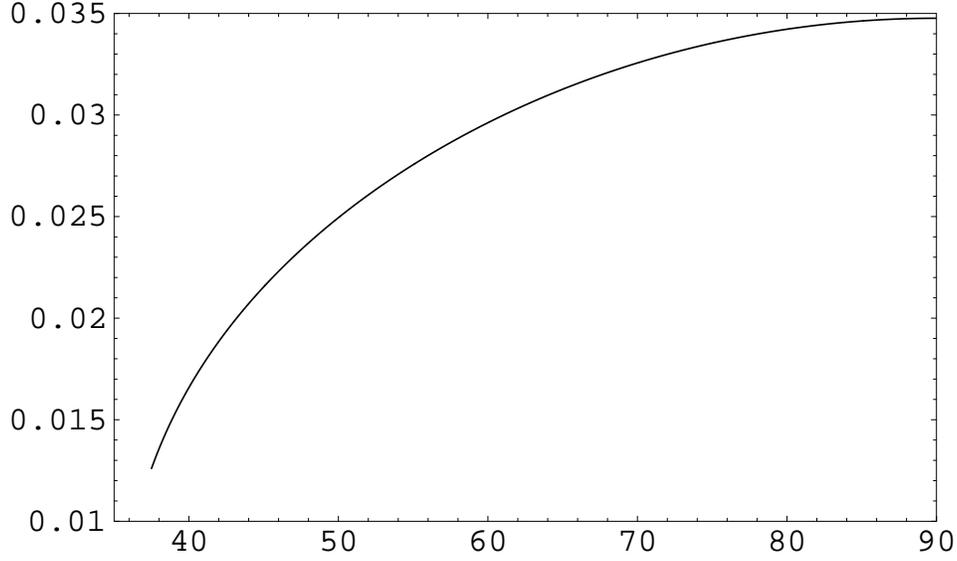, width=30pc}}
\caption{Plot of $GM/T_{\de}$
versus inclination angle $i$ (measured in degrees in this figure) based on
integration of equations~\eqref{swarmeqmx}--\eqref{swarmeqmz} with 
initial data $X=Z=Y=0$, $\dot X=V_0 \cos i$, $\dot Y=0$,
$\dot Z=V_0 \sin i$ and $r_0=100\, GM$ at $T=0$,
where $V_0=\sqrt{\Gamma_0^2-1}/\Gamma_0$ with $\Gamma_0=100$.
The quantity $T_{\de}$ is defined to be the duration of
 deceleration from $\Gamma_0=100$ to $\Gamma=80$. \label{fig:3}}
\end{figure}

\begin{figure}
\centerline{\psfig{file=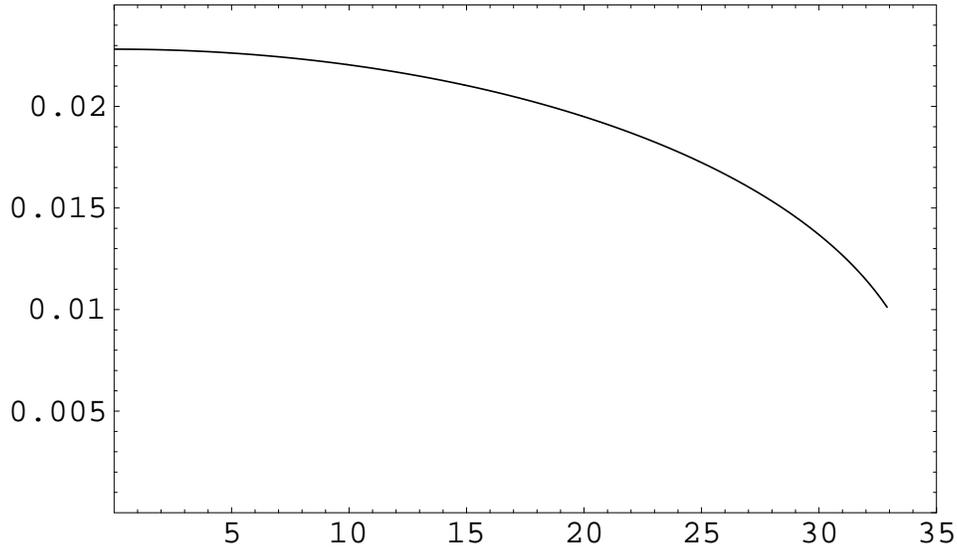, width=30pc}}
\caption{Plot of $GM/T_{\ed}$
versus inclination angle $i$ (measured in degrees in this figure) based on
integration of equations~~\eqref{swarmeqmx}--\eqref{swarmeqmz} with 
initial data $X=Z=Y=0$, $\dot X=V_0 \cos i$, $\dot Y=0$,
$\dot Z=V_0 \sin i$ and $r_0=100\, GM$ at $T=0$,
where $V_0=\sqrt{\Gamma_0^2-1}/\Gamma_0$ with $\Gamma_0=100$. \label{fig:4}}
\end{figure}

The integration of equations~\eqref{timecond} in the $(X,Z)$ 
plane shows that deceleration is maximum at $i = \pi /2$
and monotonically decreases with
decreasing inclination until about $35^\circ$
as demonstrated in Fig.~\ref{fig:3}. Moreover,
acceleration is maximum at $i = 0$ and monotonically decreases away from the $X$
axis, as demonstrated in Fig.~\ref{fig:4}, until about $35^\circ$
 when it turns into
deceleration. 
This circumstance can be illustrated by means of the critical velocity cone
in Fig.~\ref{fig:0}: a particle with velocity within the cone decelerates 
relative to
$\mathcal{O}$, while a particle with velocity outside the cone accelerates
relative to $\mathcal{O}$.

In this section, we have studied the solutions of the geodesic equation
limited to the lowest-order tidal terms
in an appropriately chosen Fermi coordinate patch. Of course, the equations
of general relativity can be expressed with respect to any admissible system
of coordinates. We choose a quasi-inertial Fermi normal coordinate system
due to its correspondence with the analysis of observational data. Inside
the Fermi coordinate patch, the geodesic equation is exact within the test
particle approximation scheme. Thus the particle velocity $\mathbf{V}$ 
can take
any value consistent with the equation of motion~\eqref{eq18} and the timelike
condition~\eqref{eq19}, see~\cite{n5}. 
The Fermi coordinate system has been employed extensively in general
relativity theory; for instance, it is used in~\cite{9} to discuss the local
bending of light in a gravitational field.
The robustness of our numerical 
results should be
emphasized. In fact, our results do not change appreciably if instead of
$\mathbf{X} ( 0 ) = 0$, the initial position of the test particle is chosen
reasonably close to the reference observer. Moreover, the reference observer
can be any relatively slow-moving test particle on a radial escape
trajectory~\cite{n6}. 
\section{Tidal Acceleration}\label{s4}
\begin{figure}
\centerline{\psfig{file=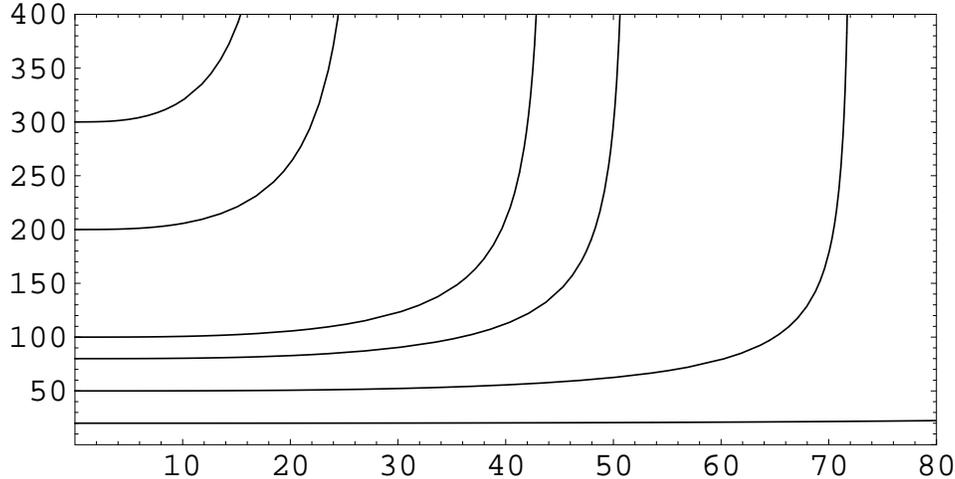, width=30pc}}
\caption{Plot of the Lorentz factor $\Gamma=1/\sqrt{1-\dot X^2+GM X^2/r^3}$
versus $T/(GM)$
based on the integration of equation~\eqref{eq33}
for $r_0=100\,GM$ and $\Gamma_0=20, 50, 80,100,200$  and 300. 
\label{fig:2}}
\end{figure}

\begin{figure}
\vspace{1in}
\centerline{\psfig{file=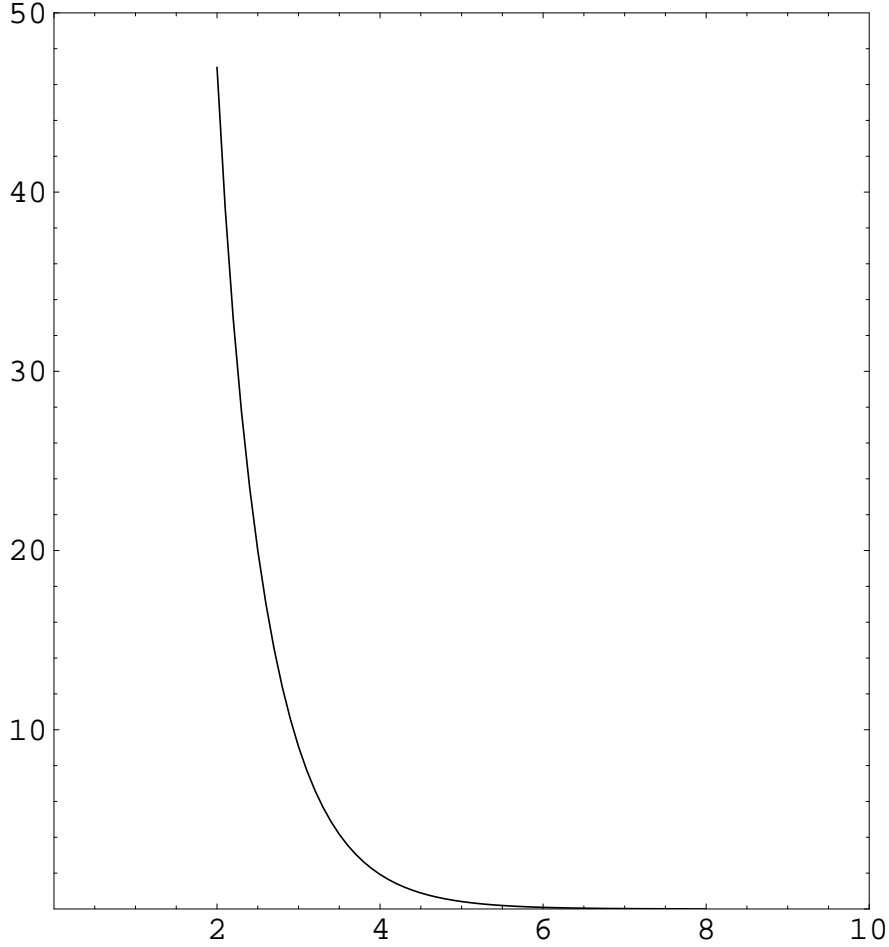, height=30pc}}
\caption{Plot of $T_{\ed}/(GM)$ versus $\log_{10}(\Gamma_0)$ based
on the integration of equation~\eqref{eq33} with $ r_0=100\, GM$,
$X=0$ and $\dot X=\sqrt{\Gamma_0^2-1}\,/\Gamma_0$ at $T=0$.  \label{fig:5}}
\end{figure}

\label{sec:3}
This section is devoted to the singular phenomenon of tidal acceleration of particles to the 
local speed of light within the Fermi coordinate system
as shown in Fig.~\ref{fig:2}. This singularity comes about due to the nonlinear character of equation \eqref{eq33}: for
$V>1/\sqrt{2}$, the nonlinear factor $1-2V^2$ changes sign thereby leading to tidal acceleration. As $\Gamma_0$
increases, $T_{\ed} /(GM)$ decreases; the graph of $T_{\ed}/(GM)$ versus $\log_{10}\Gamma_0$ is given in 
Fig.~\ref{fig:5}. It
follows from this figure that the singularity under consideration here is dynamic in origin and has nothing to do with
the kinematic breakdown of the Fermi coordinate system for 
$|\mathbf{X}| \gtrsim \mathcal{R}$.

Imagine the general solution of the geodesic equation in the Fermi
coordinate system in the $(X, Z)$ plane, since the symmetry of the
configuration permits us to set $Y(T) = 0$ for all $T$. In this case,
the generalization of
the system~\eqref{swarmeqmx}--\eqref{swarmeqmz} that takes into 
account second-order tidal terms~\cite{n8} is
given by
\begin{eqnarray}
\nonumber \ddot X+ \frac{GM}{r^3}
\big[  
X (1-2  \dot X^2-\frac{2}{3} \dot Z^2 )+\frac{14}{3} \dot X Z \dot Z \big]
&=& \frac{3 GM} {2 r^{4}} 
\big[ 2 X  Z (1-2 \dot X ^2-\frac{1}{2}\dot Z^2)\\
\label{neweq36}&&{}-(2 X ^2-5 Z^2)\dot X \dot Z\big],\\ 
\nonumber \ddot Z - \frac{2GM}{r^3}\big[ Z (1-2 \dot Z^2+\frac{1}{3}  \dot X ^2)
   +\frac{2}{3}  X  \dot X \dot Z\big]
&=&
    \frac{GM}{ 4r^{4}} \big[6  X^2-12 (1-2 \dot Z^2) Z^2\\
\label{neweq37}&&{}- 20  X\dot X Z\dot Z
 -5 Z ^2  \dot X ^2-11 X ^2  \dot Z ^2\big],
\end{eqnarray}
where we have employed the linear approximation scheme described in appendix
B of~\cite{n6} for the evaluation of the covariant derivatives of the Riemann
tensor along the reference geodesic $\mathcal{O}$. 
In fact, the relevant nonzero
covariant derivatives that are needed in the derivation of 
equations~\eqref{neweq36} 
and~\eqref{neweq37} are given by
\begin{equation}\label{neweq38} 
\fr_{1313,3} = - \fr_{0101,3} = - \fr_{0103,1} = \frac{1}{2} 
\fr_{0303,3} = 3\,\frac{GM}{r^4},
\end{equation}
except for the symmetries of the curvature tensor. The timelike condition in
this case takes the form
\begin{eqnarray}
\nonumber \frac{1}{\Gamma^2}&=& 1-\dot X^2-\dot Z^2+
      \frac{GM}{r^3}\big[X ^2-2 Z ^2-\frac{1}{3} ( X  \dot Z - Z  \dot X )^2\big]\\
\label{newgamma}&&{}- \frac{GM}{r^4}\big[(3 X^2-2 Z^2) Z 
  -\frac{1}{2}  (X \dot Z +Z  \dot X ) \dot X Z^2 \big].
\end{eqnarray}
Moreover, $\hat \Gamma = \sqrt{ -g_{00}}\, \Gamma$ in this approximation, where
\begin{eqnarray}
-g_{00}&=& 1+\frac{GM}{r^3}(X^2-2 Z^2)-\frac{GM}{r^4}(3 X^2-2 Z^2)Z.
\end{eqnarray}

We integrate equations~\eqref{neweq36} and~\eqref{neweq37} 
with initial conditions that $X = Z= 0$ at $T = 0$
 with $\dot X = V_0$ and $\dot Z = 0$; then, we use the results to
plot the behavior of $\Gamma$ and $\hat\Gamma$ versus $T/ (GM)$.
These plots turn out to be indistinguishable from figure~\ref{fig:2}, since
the contribution of the second-order tidal terms is very small; in fact,
for $\Gamma_0=50$, $T_{\ed}/(GM)$ turns out to be 72.14 for $\Gamma$ and 
$\hat\Gamma$
 compared with 72.12
in figure~\ref{fig:2}. Thus, 
the singularity in 
$\Gamma$ is moderated by the presence of the higher-order terms.

These results follow from the integration of equations that are not the
actual geodesic equation; indeed, the geodesic equation would contain the
infinite set of higher-order tidal terms, whereas we have considered only
the first and the second terms of this infinite set. Though the influence of
each higher-order term by itself may be small, as we have just demonstrated
in the case of the second-order terms, we cannot conclude that the same is
necessarily true of the \emph{whole} set. 
Physically, however, our results indicate
that \emph{ with respect to the ambient medium}
around the central source, the free
particles of the swarm can undergo significant but finite tidal
accelerations and decelerations depending upon their directions of
propagation. The observational aspects of these results are discussed in the
following section.
 
\section{Discussion}
In our work on the acceleration/deceleration phenomena, we have adopted an
approach based on \emph{relative} motion within a quasi-inertial reference frame
(i.e.\ a Fermi normal coordinate system); in fact, this treatment is 
generally consistent with the analysis of
observational data. Moreover, we have concentrated on
geodesic, that is, force-free, flow for the swarm of particles. We have
neglected plasma effects in this paper for the sake of simplicity; however,
a more thorough investigation should certainly take these into 
account~\cite{8,G}. Indeed, in the comparison of our theoretical results with observation, the
astrophysical environment around the central source should play an important
role. 

It should be emphasized that the acceleration/deceleration phenomena that
we have discussed occur within the Fermi coordinate system and are thus
measurable relative to the ambient medium surrounding the central source.
Let us note that at $T = 0$ and $r=r_0$, where the worldlines of 
$\mathcal{O}$ and $\mathcal{A}$ intersect,
\begin{equation}\label{interc} \lambda^\mu_{\;\;(0)}U_\mu = -\Gamma_0, \qquad 
 \lambda^\mu _{\;\;(3)}U_\mu = \Gamma_0 V_0 \sin i.
\end{equation}
These invariants can also be computed in the background Schwarzschild
coordinate system and we find after a straightforward calculation that
\begin{equation}\label{EJ}
 \mathcal{E}=( 1 + \sqrt{\frac{2 GM}{r_0}}\, V_0 \sin i)\Gamma_0 ,\qquad
 \mathcal{L} = r_0 \Gamma_0 V_0\cos i,
\end{equation}
where $ \mathcal{E}$ and $\mathcal{L}$ are respectively the energy and orbital angular momentum per
unit mass of the particle $\mathcal{A}$ as determined by the static inertial
observers at infinity in the exterior Schwarzschild spacetime. 
In fact, $\mathcal{E}$
and $\mathcal{L}$ are constants along the geodesic orbit of $\mathcal{A}$. Thus if this
geodesic orbit actually escapes to spatial infinity ($ \mathcal{E} \ge 1 $), it would
not appear to have any remarkable features as determined by the
asymptotically static inertial observers; for example,
for the accelerating particle discussed in section~\ref{sec:3},
 $\mathcal{E}=\Gamma_0$ and $\mathcal{L}=r_0\Gamma_0 V_0$ since $i=0$. 
On the other hand, if the motion
of $\mathcal{A}$ is referred to the ambient medium surrounding the central
source, the acceleration/deceleration phenomena that we have discussed can
be measured. This is the key point: the gain or loss of gravitational tidal
energy is measurable \emph{relative to the ambient medium}. In fact, the interaction of (charged) particles in the swarm
with the ambient medium can lead to the radiation of tidal energy to distant observers.
For instance, the collisions of such accelerated particles
with those of the ambient medium can transfer their tidal energies to the latter
particles that could then escape from the system and
appear far away as highly energetic cosmic rays.

The acceleration phenomena are consistent with the 
recent observations---by the \emph{Chandra} X-Ray Observatory~\cite{new12}---of accelerated 
motion normal to the
jet directions in four neutron stars in our galaxy: Crab Pulsar, Vela
Pulsar, PSR B1509-58 and SNR G54.1+0.3. 
A detailed analysis of the recent Crab nebula X-ray data is contained 
in~\cite{new13, new14}. The deceleration phenomena are
consistent with observations of the speed of the plasma clumps in
microquasar jets~\cite{F}. Our theoretical assumptions rely upon the
analysis of such observations based on the detection of relative motion
using the standard flat geometry of an inertial system of coordinates.
Indeed, the acceleration of particles normal to the jet direction is
measured relative to the fixed central features associated with the jets
near the source~\cite{new12}. Moreover, the 
motion of a clump within a
jet is
measured relative to certain ``fixed" features of the ambient medium~\cite{F}.

   What is the source of energy for the gravitational acceleration of
particles? While a local description of gravitational energy is not possible
in accordance with Einstein's principle of equivalence, we must nevertheless
account for the energy of the swarm of particles as measured by
the ambient medium
around the source.
Ultrarelativistic particles moving away from the source within the critical
velocity cone \emph{lose} energy and  approach the critical speed
regardless of their initial ultrarelativistic speed. 
On the other hand,
ultrarelativistic particles moving outside the critical 
velocity cone \emph{gain}
energy and accelerate as they move outward. We expect that there is
essentially a balance between the energies that are lost and gained in the
deceleration and acceleration processes, respectively.
If the
net energy that leaves the system is positive, one may speculate that the
whole system of particles around the source would slightly shrink. Such a
contraction would lead to the release of gravitational energy that could
account for the net loss of energy to the highly energetic particles that leave
the system; however, the detailed dynamics as well as the limitations of
such processes is beyond the scope of this work.

   The rotation of the central source has been neglected in this paper;
however, detailed studies have revealed that---other than specifying the
jet directions---the influence of the rotation of the central source on
our specific acceleration/deceleration results is rather 
small~\cite{5,6,n4,n5,n6}.

   Finally, we should mention that UHE cosmic rays of energy $\sim 10^{20}$ eV are
not expected to reach the Earth from extragalactic sources due to the GZK
effect~\cite{13,14,15d}. Our acceleration results indicate that the observed UHE
cosmic rays may come from microquasars or neutron stars in our galaxy. This
idea could be tested with the Pierre Auger Observatory~\cite{20}.

\appendix
\section{}
The purpose of this appendix is to study the behavior of $T_{\ed}$
as $\Gamma_0\to \infty$; indeed, we will show that in this limit 
$T_{\ed}\propto \Gamma_0^{-2/3}$.
As in the numerical work reported in this paper, 
 we will consider the solution $X(T,V_0)$ of the differential
equation
\begin{equation}\label{eq36}
\ddot{X}=-\frac{GM}{r^3} X(1-2\dot{X}^2)
\end{equation} 
with initial conditions
$X(0,V_0)=0$ and $ V(0,V_0)=V_0$,
and  
\begin{equation}  
\label{eq37}\frac{1}{\Gamma^2}=1-V^2+\frac{GM}{r^3}X^2.
\end{equation}
We note that the solution of the differential equation
has continuous partial derivatives of all orders 
with respect to $T$ and $V_0$ 
by standard results in the theory of ordinary differential 
equations (see~\cite{ccc}).

Since $\Gamma\to \infty$ if and only if the right-hand side of
equation~\eqref{eq37} vanishes,  we define 
\begin{equation}\label{eq38}
\Upsilon(T, V_0):=1-V(T,V_0)^2
+\frac{GM}{r(T)^3}X(T,V_0)^2
\end{equation}
and discuss the equation
$\Upsilon(T, V_0)=0$.

Let us observe that $\Upsilon(0, 1)=0$. That is, $\Gamma=\infty$ for
a particle whose initial velocity at $X=0$
is the speed of light. We will show that there is a \emph{unique}
implicit solution
$V_0(T)$ (i.e. $\Upsilon(T, V_0(T))=0$) such that $V_0(0)=1$ and 
$V_0(T)<1$ for $T$ near $T=0$. Moreover, for $T>0$ and near $T=0$, the implicit
solution can be inverted so that $\Upsilon(T(V_0), V_0)=0$; that is,
for each extremely ultrarelativistic solution, there is a finite time---previously
denoted as $T_{\ed}$---depending
on the initial velocity $V_0$ such that $\Gamma\to \infty$. Also, $T(V_0)$
is a decreasing function of $V_0$ such that $T(V_0)\to 0$ as $V_0\to 1$. 

The proof of these statements is simply an application of the implicit function 
theorem. 
Note that 
$\Upsilon(0, 1)=0$ and $ \Upsilon_{V_0}(0, 1)=-2$;
therefore, by the implicit function theorem, there is a function
$V_0(T)$ defined for $T$ near $T=0$ such that 
$\Upsilon(T, V_0(T))=0$ and if $\Upsilon(T^*, V_0^*)=0$ for  
$(T^*, V_0^*)$ near $(0,1)$, then $V_0(T^*)=V_0^*$.
To demonstrate that for $T>0$ 
the curve $(T, V_0(T))$ lies in the physical region ($V_0(T)<1$), we will
show that the Taylor series of $V_0$ centered at $T=0$ is given by
\begin{equation}\label{eq39}
V_0(T)=1-\frac{1}{\sqrt{2}}(GM)^{3/2} r_0^{-9/2}\, T^3+O(T^4).
\end{equation}
In fact, the remaining claims are all simple consequences of this result. 

To establish equation~\eqref{eq39}, we first note that
\begin{equation}\label{eq40}
r(T)=(r_0^{3/2}+\frac{3}{2}\sqrt{2 GM}\, T)^{2/3}
\end{equation}
and
$\ddot X(0,1)=0$. A simple computation yields
$\Upsilon_T(0,1)=0$,
and,  from  differentiating $\Upsilon ( T, V_0 (T) ) = 0$, that is,
\begin{equation}\label{eq41}
\Upsilon_T(T,V_0(T)) +\Upsilon_{V_0}(T,V_0(T)) V_0'(T)=0,
\end{equation}
we conclude that $V_0'(0)=0$, where the prime denotes differentiation
of the function $V_0$. Proceeding in the same manner
to compute the higher-order derivatives of $V_0(T)$, we find that 
$V_0''(0)=0$ and then differentiating equation~\eqref{eq41} twice leads
to 
\begin{equation}\label{eq42}
V_0'''(0)=\frac{1}{2} \Upsilon_{TTT}(0,1).
\end{equation}
Next, differentiating equation~\eqref{eq36} twice with respect to time
leads to 
\begin{equation}\label{eq43}
X_{TTT}(0,1)=\frac{GM}{r_0^3},\qquad X_{TTTT}(0,1)=2 W,
\end{equation}
where $W$ is given by
\begin{equation}\label{eq44}
W= \frac{d}{dT}\Big(\frac{GM}{r^3}\Big)\Big|_{T=0}
=-3\sqrt{2}\, \Big(\frac{GM}{r_0^3}\Big)^{3/2}.
\end{equation}
Moreover, it follows from differentiating equation~\eqref{eq38} three
times with respect to $T$ that 
\begin{equation}\label{eq45}
\Upsilon_{TT}(0,1)=0,\qquad \Upsilon_{TTT}(0,1)=-2 X_{TTTT}(0,1)+6 W.
\end{equation}
Therefore, $X_{TTTT}(0,1)=\Upsilon_{TTT}(0,1)=2 W$ and
$V'''(0)=W$,
as required.

Starting from equation~\eqref{eq39}, we can now derive a simple
approximate relationship between $\Gamma_0$ and $T_{\ed}/(GM)$
that is valid for extremely ultrarelativistic particles: 
\begin{equation}\label{eq46}
\Big(\frac{T_{\ed}}{GM}\Big)^3\approx \frac{1}{\sqrt{2}}\, 
 \Big( \frac {GM}{r_0}\Big)^{-9/2} \Gamma_0^{-2}.
\end{equation}
Thus for $r_0=100\, GM$ and $\Gamma_0=300$ we find 
$T_{\ed}/(GM)\approx 20$, in agreement with the numerical
results of Fig.~\ref{fig:5}.


\newpage
\begin{thebibliography}{xxxxx}

\bibitem{2} J.~L. Synge, {\it Relativity: The General Theory}
(North-Holland, Amsterdam, 1960).

\bibitem{5} C. Chicone and B. Mashhoon, 
{\it Class. Quantum Grav.} {\bf 19}, 4231 (2002).

\bibitem{6} 
C. Chicone, B. Mashhoon and  B. Punsly,
\emph{Int. J. Mod. Phys. D} {\bf 13}, 945 (2004).

\bibitem{n4} C. Chicone and B. Mashhoon, 
{\it Class. Quantum Grav.} {\bf 21}, L139 (2004).

\bibitem{n5} C. Chicone and B. Mashhoon, 
{\it Class. Quantum Grav.} {\bf 22}, 195 (2005).

\bibitem{n6} C. Chicone and B. Mashhoon, 
{\it Ann. Phys. (Leipzig)}  {\bf 14},  290 (2005).

\bibitem{7}
J.~D. Finley III, \emph{J. Math. Phys.} {\bf 12}, 32 (1971).

\bibitem{n8} B. Mashhoon, \emph{Astrophys. J.} {\bf 216}, 591 (1977).

\bibitem{9} J. Ehlers and W. Rindler, \emph{Gen. Rel. Grav.} {\bf 29},
519 (1997).

\bibitem{ccc} C. Chicone, 
\emph{Ordinary Differential Equations with Applications} 
(Springer, New York, 1999). 

\bibitem{8} B. Punsly, 
{\it Black  Hole Gravitohydromagnetics} (Springer-Verlag, New York, 2001).

\bibitem{G} A. W. Guthmann, M. Georganopoulos, A. Marcowith and 
K. Manolaku, eds.,
Relativistic Flows in Astrophysics, \emph{Lect. Notes Phys.} 
{\bf 589} (Springer, Berlin, 2002).
\bibitem{new12} $\langle$http://chandra.harvard.edu/photo/chronological.html$\rangle$.
\bibitem{new13} S.\ Shibata, H.\ Tomatsuri, M.\ Shimanuki, K.\ Saito and
K.\ Mori, \emph{Mon.\ Not.\  R.\  Astron.\  Soc.\ } {\bf 346}, 841 (2003).
\bibitem{new14} K.\ Mori, D.~N.\ Burrows, J.~J.\ Hester, G.~G.\ Pavlov, 
 S.\ Shibata and
H.\ Tsunemi, \emph{Astrophys.\ J. } {\bf 609}, 186 (2004).
\bibitem{F} R. Fender, in  \emph{Compact Stellar X-ray Sources}, eds. 
W.~H.~G. Lewin and M. van der Klis (Cambridge University Press,
Cambridge,  2004). 

\bibitem{13} K. Greisen, \emph{Phys. Rev. Lett.} {\bf 16}, 748 (1966).

\bibitem{14} G.~T. Zatsepin and V.~A. Kuzmin, \emph{ JETP Lett.}  {\bf 4}, 
  78 ( 1966) [\emph{Pisma Zh. Eksp.
Teor. Fiz.} {\bf  4}, 114 (1966)].

\bibitem{15d} D.~ F. Torres and L.~A. Anchordoqui, \emph{Rep. Prog. Phys.} {\bf 67},
1663 (2004). 
\bibitem{20} $\langle$http://www.auger.org$\rangle$.

\end{thebibliography}
\end{document}